\documentstyle[a4wide,epsf,11pt,titlepage]{article}

\newcounter{nref}
\newcommand{\bbib}{%
  \renewcommand{\refname}{\large\bf References}%
  \begin{thebibliography}{99}%
  \setcounter{enumiv}{\thenref}}
\newcommand{\ebib}{%
  \end{thebibliography}%
  \setcounter{nref}{\arabic{enumiv}}}
\newcommand{\head}[3]{%
  \setcounter{nref}{0}%
  \thispagestyle{empty}%
  \section*{\LARGE\bf #1}%
  \stepcounter{section}%
  \addcontentsline{toc}{section}{#1}%
  \large\itshape%
  #2\\\vspace{0.1pt}\\%
  #3%
  \normalsize\upshape%
  \bigskip}

\begin{document}


\head{Coalescing Neutron Stars: A Solution to the R-Process Problem ?}
     {S.\ Rosswog$^1$, F.K.\ Thielemann$^1$, 
      M.B.\ Davies$^2$, W.\ Benz$^3$, T.\ Piran$^4$}
     {$^1$ Departement f\"ur Physik und Astronomie, Universit\"at Basel, 
            Switzerland\\
      $^2$ Institute of Astronomy, University of Cambridge, UK\\
      $^3$  Physikalisches Institut, Universit\"at  Bern, Switzerland\\
      $^4$ Racah Institute for Physics, Hebrew University, Jerusalem, 
		Israel}

\subsection{Introduction}

Most recent nucleosynthesis parameter studies 
\cite{freiburghaus97,freiburghaus98,meyer97} place questions on the ability 
of high entropy 
neutrino wind scenarios in type II supernovae to produce r-process nuclei for 
$A < 110$ in correct amounts. In addition, it remains an open question whether
 the  entropies required for the nuclei with $A>110$ can actually be attained 
in type II supernova events. Thus, an alternative or supplementary r-process 
environment is needed, leading possibly to two different production sites
for r-process nuclei: a high entropy, high $Y_{e}$ (neutrino wind in type II 
supernovae) and a low entropy, low $Y_{e}$ (decompression of neutron star (ns)
material) scenario.\\
Further indications for a production site possibly different from SN II arise 
from observations of low metallicity stars \cite{mathews92}. It seems that the
 production of 
r-process nuclei is delayed in comparison with the major SN II yields, a fact 
that would be consistent with the merging scenario of two neutron stars.\\
The tidal disruption of a ns by a black hole and possible consequences for nucleosynthesis has first been studied by Lattimer and Schramm \cite{lattimer74,lattimer76}, the merging of a neutron star binary has been discussed by Symbalisty and Schramm \cite{symbalisty82}. The related decompression of the neutron star material has been investigated by Lattimer et al. \cite{lattimer77}, Eichler et al. \cite{eichler89}, who also discussed various other aspects of such a merging scenario, and by Meyer \cite{meyer89}. In the context of numerical simulations the merging event nucleosynthesis has been discussed by Davies et al. \cite{davies94} and by Ruffert et al. \cite{ruffert97}.

\subsection{The Calculations}
To investigate the possible relevance of neutron star mergers for the r-process
 nucleosynthesis we perform 3D Newtonian SPH calculations of the hydrodynamics
 of equal (1.6 M$_{\odot}$ of baryonic) mass  neutron star binary coalescences.
Starting with an initial separation of 45 km we follow the evolution of matter
 for 12.9 ms.
 We use the physical equation 
of state of Lattimer and Swesty \cite{lattimer91} to model the microphysics of
 the hot neutron star matter. To test the sensitivity of our results  to the 
chosen approaches and approximations we perform in total 10 different runs 
where we test each time the sensitivity to one property of our model 
\cite{rosswog98}. We vary the resolution ($\sim 21000$ and $\sim 50000$ 
particles), the equation of state (polytrope), the artificial viscosity scheme 
\cite{morris97},
 the stellar masses (1.4 M$_{\odot}$ of baryonic matter), we include neutrinos
 (free-streaming limit), switch off 
the gravitational backreaction force, and vary the initial stellar spins. In 
addition we test the influence of the initial configuration, i.e. spherical 
stars versus corotating equilibrium configurations. 

\subsection{Results}
\begin{figure}[ht]
\centerline{\epsfxsize=0.8\textwidth\epsffile{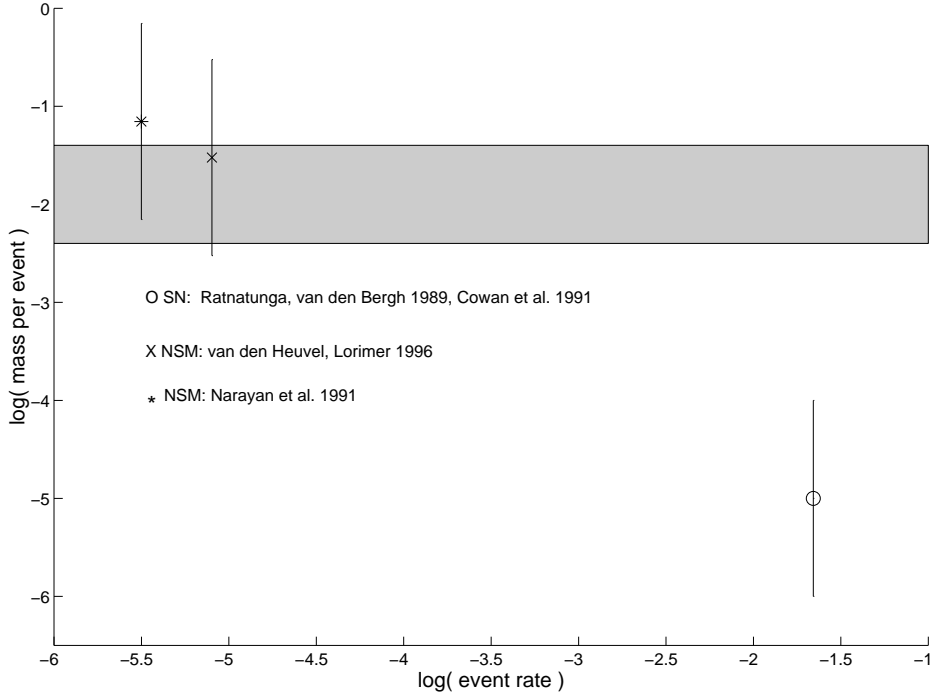}}
\caption{The shaded region shows the amount of ejecta 
found in our calculations. The circle shows the amount of ejecta needed per 
event if SN II are assumed to be the only sources of the r-process. The 
asterisk gives the 
needed ejecta per merging event for the rate of Narayan et al. (1991), the 
cross for the 
estimate of van den Heuvel and Lorimer (1996). The event rate is given in 
year$^{-1}$ galaxy$^{-1}$, the ejected mass in solar units.}
\label{rosswog.fig1}
\end{figure}

We find that, dependent on the initial spins and strongly dependent on the EOS,
 between $4\cdot10^{-3}$ and $4\cdot10^{-2}$ M$_{\odot}$ become unbound. 
Assuming a core collapse supernova rate of $2.2 \cdot 10^{-2}$ (year galaxy)
$^{-1}$ \cite{ratnatunga89}, one needs $10^{-6}$ to $10^{-4} \; {\rm M}_{\odot}
 $ of ejected r-process material per supernova event to explain the observed 
abundances if type II supernovae are assumed to be the only source. The rate 
of neutron star mergers  has recently been 
estimated to be $8 \cdot 10^{-6}$ (year galaxy)$^{-1}$ (see 
\cite{vandenheuvel96}). Taking these numbers, one would hence need $\sim 3 
\cdot 10^{-3} \; {\rm M}_{\odot}$ to $\sim 0.3 \; {\rm M}_{\odot}$ for an 
explanation of the observed r-process material  exclusively by neutron star 
mergers. Thus our  results for the ejected mass from $4 \cdot 10^{-3}$ to $3-4
 \cdot 10^{-2}  \; {\rm M}_{\odot}$ look 
very promising (see Figure \ref{rosswog.fig1}).\\

\begin{figure}[th]
 \centerline{\epsfxsize=0.6\textwidth\epsffile{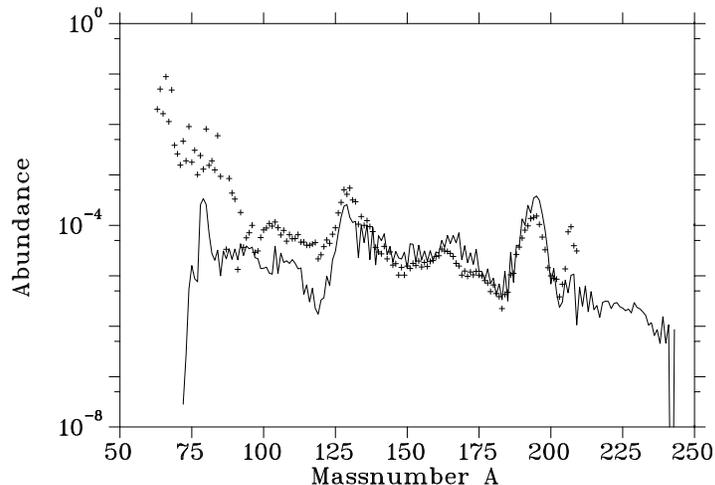}}
\caption{Comparison of the r-process calculations for a 
corotating system (initial corotation; line) with the observed abundances 
(crosses).}
\label{rosswog.fig2}
\end{figure}

As a first step we use the mean properties of all ejected particles (initial 
corotation) for an r-process calculation. We adopt the following approach: in 
the very first expansion phase (where $\rho > \rho_{drip} \approx 4\cdot 
10^{11}$ g cm$^{-3}$) we use the abundances of neutrons, protons, alphas and a
 representative nucleus provided by the LS-EOS. When the density drops below 
$\rho_{drip}$ we switch over to a treatment of individual nuclei with a full 
r-process network following all reactions like neutron capture, 
photo-disintegrations, $\beta$-decays etc. as discussed in Freiburghaus et al.
\cite{freiburghaus98}. Since the representative nucleus at $\rho_{drip}$ was 
too neutron rich ($(Z,A)=(26,155)$), we took the most neutron rich nucleus in 
the network ($(Z,A)=(26,73)$) and assumed the remaining neutrons to be free.
Following the trajectory given by the hydro calculation (extrapolation for $t 
> 12.9$ ms) we obtained the abundance pattern that is shown in Figure 
\ref{rosswog.fig2} together with the observed r-process abundances.\\

The observed features of the abundance pattern in the range $125 < A < 200$ are
 well reproduced. Especially the peak around $A=195$ is easily reproduced 
without any tuning of the initial entropy.\\
This approach has two shortcomings: (i) as long as the LS-EOS is used, only one
 (representative) nucleus is used instead of an ensemble of nuclei and (ii) 
weak interactions such as $\beta$-decays or $e^{-}-,e^{+}-$captures on protons
 and neutrons are disregarded in this early phase. For the case of initial 
corotation this approximation might not be crucial since the ejecta essentially
 stay cold (until perhaps, at later times, heating by $\beta$-decays sets in).
 For different initial spins, however, weak interactions might change the 
$Y_{e}$ of the composition in this early phase.\\
Clearly, in future investigations these aspects have to be treated in more 
detail.

\subsection*{Acknowledgements}

We thank Ch. Freiburghaus for providing us with Fig.2. S.R. thanks P. 
H\"oflich and H.T. Janka for useful discussions.


\end{document}